\documentclass[preprint,showpacs,preprintnumbers,amsmath,amssymb]{revtex4}
\usepackage{graphics}
\usepackage{dcolumn}
\usepackage{bm}

\begin{document}

\title{Unified theory for Goos-H\"{a}nchen and Imbert-Fedorov effects}

\author{Chun-Fang Li}

\affiliation{Department of Physics, Shanghai University, Shanghai 200444, P. R. China}
\affiliation{State Key Laboratory of Transient Optics and Photonics, Xi'an Institute of Optics
and Precision Mechanics of CAS, Xi'an 710119, P. R. China}

\date{\today}

\begin{abstract}
A unified theory is advanced to describe both the lateral Goos-H\"{a}nchen (GH) effect and the
transverse Imbert-Fedorov (IF) effect, through representing the vector angular spectrum of a
3-dimensional light beam in terms of a 2-form angular spectrum consisting of its 2 orthogonal
polarized components. From this theory, the quantization characteristics of the GH and IF
displacements are obtained, and the Artmann formula for the GH displacement is derived. It is
found that the eigenstates of the GH displacement are the 2 orthogonal linear polarizations in
this 2-form representation, and the eigenstates of the IF displacement are the 2 orthogonal
circular polarizations. The theoretical predictions are found to be in agreement with recent
experimental results.
\end{abstract}

\maketitle


\section{Introduction}

In 1947, Goos and H\"{a}nchen \cite{Goos-H} experimentally demonstrated that a totally
reflected light beam at a plane dielectric interface is laterally displaced in the incidence
plane from the position predicted by geometrical reflection. Artmann \cite{Artmann} in the
next year advanced a formula for this displacement on the basis of a stationary-phase
argument. This phenomenon is now referred to as Goos-H\"{a}nchen (GH) effect. In 1955, Fedorov
\cite{Fedorov} expected a transverse displacement of a totally reflected beam from the fact
that an elliptical polarization of the incident beam entails a non-vanishing transverse energy
flux inside the evanescent wave. Imbert \cite{Imbert} calculated this displacement using an
energy flux argument developed by Renard \cite{Renard} for the GH effect and experimentally
measured it. This phenomenon is usually called Imbert-Fedorov (IF) effect. The investigation
of the GH effect has been extended to the cases of partial reflection and transmission in
transmitting configurations \cite{Hsue-T, Li-W} and to other areas of physics, such as
acoustics \cite{Briers-LS}, nonlinear optics \cite{Jost-AS}, plasma physics \cite{Yin-HLFZ},
and quantum mechanics \cite{Renard, Ignatovich}. And the IF effect has been connected with the
angular momentum conservation and the Hall effect of light \cite{Onoda-MN, Bliokh-B}. But the
comment of Beauregard and Imbert \cite{Beauregard-I} is still valid up to now that there are,
strictly speaking, no completely rigorous calculations of the GH or IF displacement. Though
the argument of stationary phase was used to explain \cite{Artmann} the GH displacement and to
calculate the IF displacement \cite{Schilling}, it was on the basis of the formal properties
of the Poynting vector inside the evanescent wave \cite{Beauregard-I} that the quantization
characteristics were acquired for both the GH and IF displacements in total reflection. On the
other hand, it has been found that the GH displacement in transmitting configurations has
nothing to do with the evanescent wave \cite{Li-W}.

The purpose of this paper is to advance a unified theory for the GH and IF effects through
representing the vector angular spectrum of a 3-dimensional (3D) light beam in terms of a
2-form angular spectrum, consisting of its 2 orthogonal polarizations. From this theory, the
quantization characteristics of the GH and IF displacements are obtained, and the Artmann
formula \cite{Artmann} for the GH displacement is derived. The amplitude of the 2-form angular
spectrum describes the polarization state of a beam in such a way that the eigenstates of the
GH displacement are the 2 orthogonal linear polarizations and the eigenstates of the IF
displacement are the 2 orthogonal circular polarizations.

\section{General theory}
Consider a monochromatic 3D light beam in a homogenous and isotropic medium of refractive
index $n$ that intersects the plane $x=0$. In order to have a beam representation that can
describe the propagation parallel to the $x$-axis, the vector electric field of the beam is
expressed in terms of its vector angular spectrum as follows \cite{Ghatak-T},
\begin{equation} \label{field-distribution}
\mathbf{E}(\mathbf{r})= \frac{1}{2\pi} \int^{\infty}_{-\infty} \int^{\infty}_{-\infty}
\mathbf{A} (k_y,k_z) \exp(i \mathbf{k} \cdot \mathbf{r})d k_y d k_z,
\end{equation}
where time dependence $\exp(-i \omega t)$ is assumed and suppressed,
$\mathbf{A}=(\begin{array}{ccc} A_x&A_y&A_z \end{array})^T$ is the vector amplitude of the
angular spectrum, $\mathbf{k}= (\begin{array}{ccc} k_x&k_y&k_z \end{array})^T$ is the wave
vector satisfying $k^2_x+k^2_y+k^2_z=k^2$, $k=2n \pi/\lambda_0$, $\lambda_0$ is the vacuum
wavelength, superscript $T$ means transpose, and the beam is supposed to be well collimated so
that its angular distribution function is sharply peaked around the principal axis
$(k_{y0},k_{z0})$ and that the integration limits have been extended to $\pm \infty$ for both
variables $k_y$ and $k_z$ \cite{Marcuse}. When this beam intersects the plane $x=0$, the
electric field distribution on this plane is thus
$$
\mathbf{\Psi}(y,z) \equiv \mathbf{E}(\mathbf{r})|_{x=0}= \frac{1}{2\pi} \int \int \mathbf{A}
e^{i(k_y y +k_z z)} d k_y d k_z,
$$
hereafter the integration limits will be omitted as such. The position coordinates of the
centroid of the beam (\ref{field-distribution}) on the plane $x=0$ are defined by
\begin{equation} \label{y-expectation}
\langle y \rangle= \frac{\int \int\mathbf{\Psi}^{\dagger} y \mathbf{\Psi} dy dz}{\int
\int\mathbf{\Psi}^{\dagger} \mathbf{\Psi} dy dz}= \frac{\int \int\mathbf{A}^{\dagger} i
\frac{\partial \mathbf{A}}{\partial k_y} dk_y dk_z}{\int \int\mathbf{A}^{\dagger} \mathbf{A} dk_y
dk_z}
\end{equation}
and
\begin{equation} \label{z-expectation}
\langle z \rangle= \frac{\int \int\mathbf{\Psi}^{\dagger} z \mathbf{\Psi} dy dz}{\int
\int\mathbf{\Psi}^{\dagger} \mathbf{\Psi} dy dz}= \frac{\int \int\mathbf{A}^{\dagger} i
\frac{\partial \mathbf{A}}{\partial k_z} dk_y dk_z}{\int \int\mathbf{A}^{\dagger} \mathbf{A} dk_y
dk_z},
\end{equation}
where $\frac{\partial}{\partial k_y}$ means partial derivative with respect to $k_y$ with $k_z$
fixed, $\frac{\partial}{\partial k_z}$ means partial derivative with respect to $k_z$ with $k_y$
fixed, and superscript $\dagger$ stands for transpose conjugate.

Since the Fresnel formula for the amplitude reflection coefficient at a dielectric interface
depends on whether the incident plane wave is in $s$ or $p$ polarization, it is profitable to
represent the vector amplitude of the angular spectrum in terms of its $s$ and $p$ polarized
components. To this end, let us first consider one plane-wave element of the angular spectrum
whose wave vector is $\mathbf{k}^0= (\begin{array}{ccc} k \cos \theta&k \sin \theta&0
\end{array})^T$, where $\theta$ is its incidence angle. Its vector amplitude is given by
$\mathbf{A}^0=\mathbf{A}^0_s+ \mathbf{A}^0_p \equiv A_s \mathbf{s}^0+ A_p \mathbf{p}^0$, where
$A_s$ and $A_p$ are the complex amplitudes of $\mathbf{A}^0_s$ and $\mathbf{A}^0_p$,
respectively, $\mathbf{s}^0= (\begin{array}{ccc} 0&0&1 \end{array})^T$ is the unit vector of
$\mathbf{A}^0_s$ and is perpendicular to the plane $xoy$, and $\mathbf{p}^0=
(\begin{array}{ccc} -\sin \theta & \cos \theta & 0 \end{array})^T$ is the unit vector of
$\mathbf{A}^0_p$ and is parallel to the plane $xoy$. This means that $\mathbf{A}^0$ can be
represented as
$$
\mathbf{A}^0= \left( \begin{array}{cc} 0 & -\sin\theta\\0 & \cos\theta\\1 &0 \end{array}
\right) \tilde{A},
$$
where
\begin{equation} \label{two-form}
\tilde{A}=\left( \begin{array}{c} A_s\\A_p \end{array} \right) \equiv A_s \tilde{s}+A_p
\tilde{p}
\end{equation}
is what we introduce in this paper and is referred to as the 2-form amplitude of the angular
spectrum, $\tilde{s}= \left( \begin{array}{c} 1\\0 \end{array} \right)$ represents the
normalized state of s polarization, and $\tilde{p}= \left( \begin{array}{c} 0\\1 \end{array}
\right)$ represents the normalized state of p polarization. $\tilde{s}$ and $\tilde{p}$ form
the orthogonal complete set of linear polarizations.

\begin{figure}[ht]
\includegraphics{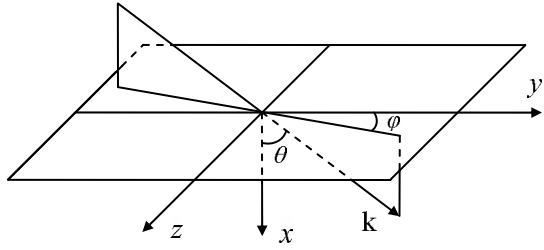}
\caption{\label{angular spectrum description} Schematic diagram for the rotation of the vector
amplitude $\mathbf{A}$ of an arbitrarily polarized light beam.}
\end{figure}
After this element is rotated by angle $\varphi$ around the $x$-axis as is displayed in Fig.
\ref{angular spectrum description}, its wave vector becomes
$$\mathbf{k}= \mathrm{M}(\varphi) \mathbf{k}^0=
\left(
      \begin{array}{c}
         k \cos\theta\\
         k \sin\theta \cos\varphi\\
         k \sin\theta \sin\varphi
      \end{array}
\right),
$$
and its vector amplitude becomes
\begin{equation}
\label{vector form of angular spectrum} \mathbf{A}
  =\mathrm{M}(\varphi) \mathbf{A}^0
  =A_s \mathbf{s}+ A_p \mathbf{p},
\end{equation}
where
$$
\mathrm{M}(\varphi)= \left(
      \begin{array}{ccc}
      1 & 0 & 0\\
      0 & \cos \varphi & -\sin \varphi\\
      0 & \sin \varphi &  \cos \varphi
      \end{array}
\right)
$$
is the rotation matrix,
$$
\mathbf{s}= \mathrm{M}(\varphi) \mathbf{s}^0= (\begin{array}{ccc} 0 & -\sin \varphi & \cos
\varphi \end{array})^T
$$
is the unit vector of $\mathbf{A}_s= \mathrm{M}(\varphi) \mathbf{A}^0_s$, and
$$
\mathbf{p}= \mathrm{M}(\varphi) \mathbf{p}^0= \left( \begin{array}{c} -\sin \theta \\ \cos\theta \cos\varphi \\
\cos\theta \sin\varphi \end{array} \right)
$$
is the unit vector of $\mathbf{A}_p= \mathrm{M}(\varphi) \mathbf{A}^0_p$. This shows that the
vector amplitude (\ref{vector form of angular spectrum}) can be represented as
\begin{equation} \label{angular spectrum in 2-form}
\mathbf{A}
  =\left( \begin{array}{cc} 0 & p_x\\s_y & p_y\\s_z & p_z \end{array} \right) \tilde{A}
  \equiv \mathrm{P} \tilde{A},
\end{equation}
where matrix $\mathrm{P}$ represents the projection of 2-form amplitude $\tilde{A}$ onto
vector amplitude $\mathbf{A}$ and is thus referred to as projection matrix, and
\begin{eqnarray}
s_y=-\sin\varphi=           -\frac{k_z}{(k^2_y+k^2_z)^{1/2}} \nonumber\\
s_z=\cos\varphi=             \frac{k_y}{(k^2_y+k^2_z)^{1/2}} \nonumber\\
p_x=-\sin\theta=            -\frac{(k^2_y+k^2_z)^{1/2}}{k} \nonumber\\
p_y=\cos\theta \cos\varphi=  \frac{k_x k_y}{k(k^2_y+k^2_z)^{1/2}} \nonumber\\
p_z=\cos\theta \sin\varphi=  \frac{k_x k_z}{k(k^2_y+k^2_z)^{1/2}} \nonumber
\end{eqnarray}

\begin{figure}[ht]
\includegraphics{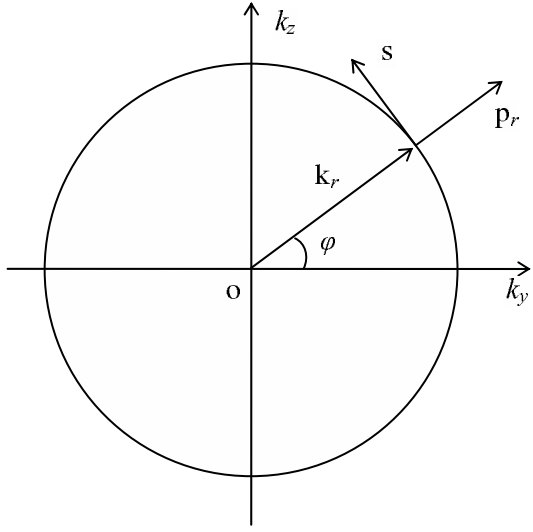}
\caption{$\mathbf{s}$ and $\mathbf{p}_r$ are in the azimuthal and radial directions,
respectively.} \label{polarization in wavevector space}
\end{figure}

Now we have successfully represented, through the projection matrix, the vector amplitude
$\mathbf{A}$ in terms of the 2-form amplitude $\tilde{A}$, that is to say, in terms of the 2
orthogonal linear polarizations $\mathbf{A}_s$ and $\mathbf{A}_p$. It should be pointed out
that in this representation, $\mathbf{s}$ is not necessarily perpendicular to the plane $xoy$,
and $\mathbf{p}$ is not necessarily parallel to this plane. Denoting $\mathbf{k}_r= k_r
\mathbf{e}_r= k_y \mathbf{e}_y+ k_z \mathbf{e}_z$, where $k_y=k_r \cos \varphi$, $k_z=k_r \sin
\varphi$, $\mathbf{e}_y$ and $\mathbf{e}_z$ are the unit vectors in $y$ and $z$ directions,
respectively, and $\mathbf{e}_r$ is the unit vector in the radial direction, we find that
$\mathbf{s}$ is in fact the unit vector in the azimuthal direction, $\mathbf{s}=
\mathbf{e}_\varphi$. Furthermore, letting $\mathbf{p}_r= p_y \mathbf{e}_y+ p_z \mathbf{e}_z$,
it is apparent that $\mathbf{p}_r= \frac{k_x}{k} \mathbf{e}_r$. In other words, $\mathbf{p}_r$
is in the radial direction. The directions of $\mathbf{s}$ and $\mathbf{p}_r$ are
schematically shown in Fig. \ref{polarization in wavevector space}.

Unit vectors $\mathbf{s}$ and $\mathbf{p}$ and the wave vector $\mathbf{k}$ are orthogonal to
each other and thus satisfy the following relations,
\begin{equation} \label{orthogonality}
\begin{array}{r}
s^2_y+s^2_z=1 \\
p^2_x+p^2_y+p^2_z=1 \\
s_y p_y+s_z p_z=0 \\
k_y s_y+ k_z s_z=0 \\
k_x p_x+ k_y p_y+ k_z p_z=0
\end{array}
\end{equation}
The first 3 equations guarantee
\begin{equation} \label{inner product of angular spectrum}
\mathbf{A}^{\dagger} \mathbf{A}= \tilde{A}^{\dagger} \tilde{A}.
\end{equation}
From expression (\ref{angular spectrum in 2-form}) for the vector amplitude and with the help
of Eq. (\ref{orthogonality}), we obtain
\begin{eqnarray}
\label{key expression1} \mathbf{A}^{\dagger} \frac{\partial \mathbf{A}}{\partial k_y}=
\tilde{A}^{\dagger} \frac{\partial \tilde{A}}{\partial k_y} -\frac{k_x k_z}{k(k^2_y+k^2_z)}
(A^*_s A_p-A^*_p A_s),\\
\label{key expression2} \mathbf{A}^{\dagger} \frac{\partial \mathbf{A}}{\partial k_z}=
\tilde{A}^{\dagger} \frac{\partial \tilde{A}}{\partial k_z} +\frac{k_x k_y}{k(k^2_y+k^2_z)}
(A^*_s A_p-A^*_p A_s).
\end{eqnarray}
Eqs. (\ref{y-expectation}), (\ref{z-expectation}), (\ref{inner product of angular spectrum}),
(\ref{key expression1}), and (\ref{key expression2}) are the central results of this paper,
from which the GH and IF displacements are derived below.

\section{Description of incident and reflected beams}

Without loss of generality, we consider an arbitrarily polarized \textit{incident} beam of the
following 2-form amplitude,
\begin{equation} \label{incident 2-form amplitude}
\tilde{A}_i= (l_{i1} \tilde{s}+ l_{i2} \tilde{p}) A \equiv \tilde{L}_i A,
\end{equation}
where $\tilde{L}_i=l_{i1} \tilde{s}+ l_{i2} \tilde{p}$ describes the polarization state of the
beam and is assumed to satisfy the normalization condition
\begin{equation} \label{normalization1}
\tilde{L}^{\dagger}_i \tilde{L}_i=1,
\end{equation}
angular distribution function $A(k_y,k_z)$ is assumed to be a positively-definite
sharply-peaked symmetric function around the principal axis $(k_{y0},k_{z0})=(k \sin
\theta_0,0)$ and satisfy the normalization condition
\begin{equation}\label{normalization2}
\int \int A^2(k_y,k_z) dk_y dk_z=1,
\end{equation}
and $\theta_0$ stands for the incidence angle of the beam. Eqs. (\ref{normalization1}) and
(\ref{normalization2}) guarantee the following normalization condition for the 2-form
amplitude (\ref{incident 2-form amplitude}),
\begin{equation} \label{normalization}
\int \int \tilde{A}^{\dagger}_i \tilde{A}_i dk_y dk_z=1.
\end{equation}
One example of such a distribution function that satisfies normalization condition
(\ref{normalization2}) is the following Gaussian function \cite{Ghatak-T, Zhang-L},
\begin{equation} \label{Gaussian distribution}
A_G=\left( \frac{w_y w_z}{\pi} \right)^{\frac{1}{2}}
    \exp\left[ -\frac{w^2_y}{2} (k_y-k_{y0})^2 \right]
    \exp\left( -\frac{w^2_z}{2} k^2_z \right),
\end{equation}
where $w_y=w_0/\cos \theta_0$, $w_z=w_0$, $w_0$ is half the width of the beam at waist.
$\Delta \theta= \frac{1}{k w_0}$ is half the divergence angle of the beam.

According to Eq. (\ref{angular spectrum in 2-form}), the vector amplitude of the incident beam
is given by $\mathbf{A}_i=\mathrm{P} \tilde{A}_i$. For a uniformly polarized beam that was
obtained from a linearly polarized beam in experiments \cite{Pillon-GG, Pillon-GGLKG,
Pillon-GGLE}, the $s$ components of all its plane-wave elements are in the same direction, and
the same to the $p$ components. But in our representation advanced here, the $s$ polarizations
of different plane-wave elements are generally in different directions; so are the $p$
polarizations. Considering Eqs. (\ref{angular spectrum in 2-form}), (\ref{incident 2-form
amplitude}) and (\ref{Gaussian distribution}) together, one concludes that in order to
describe a uniformly polarized beam mentioned above, it is essential that the incidence angle
$\theta_0$ be much larger than $\Delta \theta$. So we will only consider the case of large
$\theta_0$ below. Fortunately, this is just what we have in the case of total reflection.

It will be convenient to express $\tilde{L}_i$ on the orthogonal complete set of circular
polarizations as follows,
\begin{equation} \label{unitary transformation}
\tilde{L}_i=c_{i1} \tilde{r}+ c_{i2} \tilde{l}=\mathrm{U} \tilde{C}_i,
\end{equation}
where $c_{i1}$ represents the complex amplitude of right circular polarization, $c_{i2}$
represents the complex amplitude of left circular polarization, $\tilde{r}=\mathrm{U}
\tilde{s}= \frac{1}{\sqrt{2}} \left( \begin{array}{c} 1\\-i
\end{array} \right)$ is the normalized state of right circular polarization,
$\tilde{l}=\mathrm{U} \tilde{p}= \frac{1}{\sqrt{2}} \left( \begin{array}{c} 1\\i \end{array}
\right)$ is the normalized state of left circular polarization, $\mathrm{U}$ is the unitary
transformation matrix $\mathrm{U}=\frac{1}{\sqrt{2}}
            \left(
                  \begin{array}{cc}
                  1 & 1\\-i & i
                  \end{array}
            \right)
$, and $\tilde{C}_i=\left( \begin{array}{c} c_{i1}\\c_{i2} \end{array} \right)$. $\tilde{r}$
and $\tilde{l}$ form the orthogonal complete set of circular polarizations. Unitary
transformation guarantees $\tilde{L}^{\dagger}_i \tilde{L}_i=\tilde{C}^{\dagger}_i
\tilde{C}_i$.

When the beam is reflected at plane $x=0$, the reflected beam has the following 2-form
amplitude,
\begin{equation} \label{reflection angular spectrum}
\tilde{A}_r=\mathrm{R}_l \tilde{A}_i=\tilde{L}_r A,
\end{equation}
where
$$
\mathrm{R}_l=\left(
                   \begin{array}{cc}
                      R_s & 0\\0 & R_p
                   \end{array}
             \right)
$$
is the reflection coefficient matrix, $\tilde{L}_r=\mathrm{R}_l \tilde{L}_i \equiv \left(
\begin{array}{c} l_{r1}\\l_{r2} \end{array} \right)$ describes the polarization state of
the reflected beam, and $R_s \equiv |R_s| \exp(i \Phi_s)$ and $R_p \equiv |R_p| \exp(i
\Phi_p)$ are the reflection coefficients for s and p polarizations, respectively. It will be
convenient to express $\tilde{L}_r$ on the orthogonal complete set of circular polarizations
as follows,
\begin{equation} \label{unitary transformation for reflected beam}
\tilde{L}_r=c_{r1} \tilde{r}+ c_{r2} \tilde{l}=\mathrm{U} \tilde{C}_r,
\end{equation}
where $c_{r1}$ represents the complex amplitude of right circular polarization for reflected
beam, $c_{r2}$ represents the complex amplitude of left circular polarization,
$\tilde{C}_r=\left( \begin{array}{c} c_{r1}\\c_{r2} \end{array} \right)=\mathrm{R}_c
\tilde{C}_i$, and $\mathrm{R}_c=\mathrm{U}^{\dagger}\mathrm{R}_l \mathrm{U}$. Unitary
transformation guarantees $\tilde{L}^{\dagger}_r \tilde{L}_r= \tilde{C}^{\dagger}_r
\tilde{C}_r$.

\section{GH effect and its quantization}

Applying Eqs. (\ref{y-expectation}), (\ref{inner product of angular spectrum}), and (\ref{key
expression1}) to $\tilde{A}_i$ produces the $y$ coordinate of the centroid of the incident
beam on the plane $x=0$,
$$
\langle y \rangle_i=0.
$$
Since $R_s$ and $R_p$ are all even functions of $k_z$, we have for the $y$ coordinate of the
centroid of the reflected beam on the plane $x=0$, on applying Eqs. (\ref{y-expectation}),
(\ref{inner product of angular spectrum}), and (\ref{key expression1}) to $\tilde{A}_r$,
\begin{equation} \label{reflection y position}
\langle y \rangle_r= -\frac{1}{\mathfrak{R}}
  \int\int \left( |l_{r1}|^2 \frac{\partial \Phi_s}{\partial k_y}
                 +|l_{r2}|^2 \frac{\partial \Phi_p}{\partial k_y}
           \right)   A^2 dk_ydk_z,
\end{equation}
where
\begin{equation} \label{reflectivity}
\mathfrak{R}=\int \int (|l_{r1}|^2+ |l_{r2}|^2) A^2 dk_y dk_z
\end{equation}
describes the reflectivity of a 3D beam. The above equation can also be written as
$$
\mathfrak{R}=|l_{i1}|^2 \mathfrak{R}_s+ |l_{i2}|^2 \mathfrak{R}_p,
$$
with
$$
\mathfrak{R}_s=\int \int |R_s|^2 A^2 dk_y dk_z
$$
and
$$
\mathfrak{R}_p=\int \int |R_p|^2 A^2 dk_y dk_z.
$$

The displacement of $\langle y \rangle_r$ from $\langle y \rangle_i$ is the GH effect and is
thus given by
\begin{equation} \label{GGH displacement}
D_{GH}=
 -\frac{|l_{i1}|^2}{\mathfrak{R}} \int\int |R_s|^2 A^2 \frac{\partial \Phi_s}{\partial
   k_y}dk_ydk_z
 -\frac{|l_{i2}|^2}{\mathfrak{R}} \int\int |R_p|^2 A^2 \frac{\partial \Phi_p}{\partial
   k_y}dk_ydk_z.
\end{equation}
It is obviously quantized with eigenstates the $s$ and $p$ polarization states. The
eigenvalues are
$$
D_{GHj}= -\frac{1}{\mathfrak{R}_j} \int\int |R_j|^2 A^2 \frac{\partial \Phi_j}{\partial k_y}
  dk_y dk_z,
$$
with $j=s,p$. When the angular distribution function $A(k_y,k_z)$ is so sharp that
$\frac{\partial \Phi_s}{\partial k_y}$ and $\frac{\partial \Phi_p}{\partial k_y}$ are
approximately constant in the area in which $A$ is appreciable, we arrive at the Artmann
formula \cite{Artmann},
\begin{equation} \label{Artmann formula}
D_{GHj}= -\frac{\partial \Phi_j}{\partial k_y}.
\end{equation}
It is now clear that the quantization description of GH displacement depends closely on the
2-form representation of the angular spectrum.

\subsection{Total reflection}

When the beam is \textit{totally} reflected, the reflection coefficients take the form of
\begin{equation}\label{total reflection coefficients}
R_s=\exp(i \Phi_s), \hspace{1cm} R_p=\exp(i \Phi_p),
\end{equation}
and $\mathfrak{R}=1$. Substituting Eq. (\ref{total reflection coefficients}) into Eq.
(\ref{GGH displacement}), we obtain
$$
D_{GH}= -\int \int \left( |l_{i1}|^2 \frac{\partial \Phi_s}{\partial k_y}
                         +|l_{i2}|^2 \frac{\partial \Phi_p}{\partial k_y}
                   \right) A^2
         dk_y dk_z.
$$
If $\frac{\partial \Phi_s}{\partial k_y}$ and $\frac{\partial \Phi_p}{\partial k_y}$ are
approximately constant in the area in which $A$ is appreciable, the reflected beam maintains
the shape of the incident beam \cite{Shi-LW} and the GH displacement takes the form of
$$
D_{GH}= -|l_{i1}|^2 \frac{\partial \Phi_s}{\partial k_y}
        -|l_{i2}|^2 \frac{\partial \Phi_p}{\partial k_y},
$$
which leads naturally to the Artmann formula (\ref{Artmann formula}) for $s$ or $p$
polarization and agrees well with the recent experimental results \cite{Pillon-GG,
Pillon-GGLKG}.

\subsection{Partial reflection and generalized GH displacement}
When the beam is partially reflected, the reflected beam is also displaced from $\langle y
\rangle_i$ to $\langle y \rangle_r$ in $y$-direction. This is the so-called generalized GH
displacement \cite{Li-W} and is given by Eq. (\ref{GGH displacement}). Such generalized GH
displacements may also occur in attenuated total reflection \cite{Yin-HLFZ}, amplified total
reflection \cite{Fan-DW}, and in reflections from absorptive \cite{Lai-C} and active
\cite{Yan-CL} media. If $\frac{\partial \Phi_s}{\partial k_y}$ and $\frac{\partial
\Phi_p}{\partial k_y}$ are approximately constant in the area in which $A(k_y,k_z)$ is
appreciable, Eq. (\ref{GGH displacement}) reduces to
$$
D_{GH}=
 -\frac{|l_{i1}|^2 \mathfrak{R}_s}{\mathfrak{R}} \frac{\partial \Phi_s}{\partial k_y}
 -\frac{|l_{i2}|^2 \mathfrak{R}_p}{\mathfrak{R}} \frac{\partial \Phi_p}{\partial k_y},
$$
which also leads to the Artmann formula (\ref{Artmann formula}) for $s$ or $p$ polarized
beams.

\section{IF effect and its quantization}

Now let us pay our attention to the problem of the IF effect. As before, we first want to find
out the $z$ coordinate of the centroid of the incident beam on the plane $x=0$. On applying
Eqs. (\ref{z-expectation}), (\ref{inner product of angular spectrum}), and (\ref{key
expression2}) to $\tilde{A}_i$ and with the help of Eq. (\ref{unitary transformation}), we
have
\begin{equation} \label{z-position}
\langle z \rangle_i= (|c_{i1}|^2-|c_{i2}|^2) \langle z \rangle^c_i,
\end{equation}
where
$$
\langle z \rangle^c_i=\frac{1}{k} \int \int \frac{k_x k_y}{k^2_y+k^2_z} A^2 dk_y dk_z.
$$
Eq. (\ref{z-position}) shows that $\langle z \rangle_i$ does not vanish and is quantized with
eigenstates the 2 circular polarizations. The eigenvalues are the same in magnitude and
opposite in direction. For the Gaussian distribution function (\ref{Gaussian distribution}),
we have \cite{comment}
$$
\langle z \rangle^c_i \approx \frac{1}{k \tan \theta_0}
$$
at large incidence angle, $\theta_0 \gg \Delta \theta$.

The non-vanishing transverse displacement of the incident beam from the plane $z=0$ is in fact
an evidence of the so-called translational inertial spin effect of light that was once
discussed by Beauregard \cite{Beauregard}. Beauregard found that although the transverse wave
vector of a 2-dimensional beam is identically zero, the 2 circular polarizations have
non-vanishing transverse Poynting vector, and called this phenomenon the translational
inertial spin effect. The problem is that the electromagnetic field of so defined
2-dimensional beam is uniform in transverse direction, $\frac{\partial}{\partial z}=0$. In
order to observe this effect, it is necessary to have a bound beam that is not transversely
uniform, provided that the expectation of transverse wave vector is zero. The 3D beam that we
consider here is such a beam satisfying
$$
\langle k_z \rangle=\int \int \mathbf{A}^{\dagger} k_z \mathbf{A} dk_y dk_z
                   =\int \int \tilde{A}^{\dagger} k_z \tilde{A} dk_y dk_z
                   =0.
$$

\begin{figure}[ht]
\includegraphics{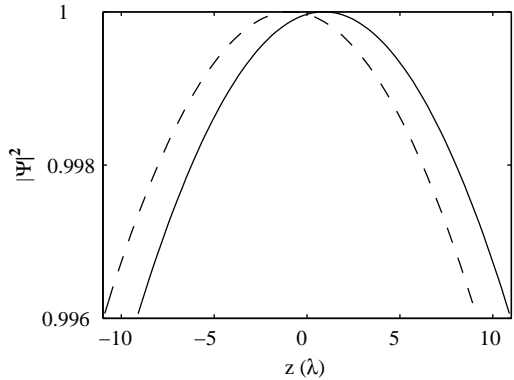}
\caption{Normalized intensity distributions of right circularly polarized beam (solid curve)
and left circularly polarized beam (dashed curve) on the $z$-axis, where $\theta_0=
10^{\circ}$, $\Delta \theta= 10^{-3}$, and $z$ is in units of $\lambda$.} \label{Tise}
\end{figure}
For example, when $\theta_0=10^{\circ}$, we have $\langle z \rangle^c_i \approx 0.9 \lambda$.
This displacement has been confirmed by the numerical calculation of the field intensity
distribution, $|\mathbf{\Psi}(0,z)|^2$, on $z$-axis as is shown in Fig. \ref{Tise}, where the
Gaussian distribution function (\ref{Gaussian distribution}) is considered with $\Delta
\theta=10^{-3} \ll \theta_0$. The right circularly polarized beam (solid curve) is displaced
$0.9\lambda$ to the positive direction, and the left circularly polarized beam (dashed curve)
is displaced $0.9\lambda$ to the negative direction.

When the beam is totally reflected, the 2-form amplitude of the reflected beam is represented
by Eq. (\ref{reflection angular spectrum}), with $R_s$ and $R_p$ given by Eq. (\ref{total
reflection coefficients}). Applying Eqs. (\ref{z-expectation}), (\ref{inner product of angular
spectrum}), and (\ref{key expression2}) to this amplitude and with the help of Eq.
(\ref{unitary transformation for reflected beam}) gives the transverse displacement of the
reflected beam from the plane $z=0$. So defined displacement is the IF effect \cite{Imbert,
Beauregard-I, Pillon-GG, Pillon-GGLKG} and is given by
\begin{equation}\label{IF effect}
D_{IF} \equiv \langle z \rangle_r=
  \frac{1}{k} \int \int (|c_{r1}|^2-|c_{r2}|^2) \frac{k_x k_y}{k^2_y+k^2_z} A^2 dk_y dk_z.
\end{equation}
This shows that the IF displacement of the reflected beam is quantized with eigenstates the 2
circular polarizations. The eigenvalues are the same in magnitude and are opposite in
direction. Eqs. (\ref{z-position}) and (\ref{IF effect}) indicate that the quantization
description of IF displacement depends closely on the 2-form representation of the angular
spectrum.

In order to compare with the recent experimental results \cite{Pillon-GG, Pillon-GGLKG,
Pillon-GGLE}, we consider such an incident beam that has the following elliptical polarization
and Gaussian distribution function,
\begin{equation}\label{ellipse}
\tilde{A}_i=\left(
                  \begin{array}{c} \cos \psi\\ e^{-i \phi} \sin \psi \end{array}
            \right)
            A_G,
\end{equation}
where $0 \leq \psi \leq \pi/2$. In this case, the IF displacement of totally reflected beam is
\begin{equation} \label{IF effect for Gaussian beam}
D_{IF}=\frac{w_y w_z \sin(2 \psi)}{k \pi}
       \int \int \frac{k_x k_y}{k^2_y+k^2_z} e^{-w^2_y (k_y-k_{y0})^2}
       e^{-w^2_z k^2_z} \sin(\phi+\Phi_s-\Phi_p) dk_y dk_z.
\end{equation}
Since Eq. (\ref{IF effect for Gaussian beam}) holds whether the beam is totally reflected by a
single dielectric interface \cite{Pillon-GG} or by a thin dielectric film in a resonance
configuration \cite{Pillon-GGLKG}, it is no wonder that the observed IF displacement in the
resonance configuration \cite{Pillon-GGLKG} is not enhanced in the way that the lateral GH
displacement is enhanced.

If the total reflection takes place at a single dielectric interface and the incidence angle
is far away from the critical angle for total reflection and the angle of grazing incidence in
comparison with $\Delta \theta$, the first and the last factors of the integrand in Eq.
(\ref{IF effect for Gaussian beam}) can be regarded as constants for a well-collimated beam
\cite{Shi-LW} and thus can be taken out of the integral with $k_y$, $k_z$, $\Phi_s$, and
$\Phi_p$ evaluated at $k_y=k_{y0}$ and $k_z=k_{z0}$, producing
\begin{equation} \label{reflection z-position for large incidence angle}
D_{IF}=\frac{\sin(2 \psi) \sin(\phi+\Phi_{s0}-\Phi_{p0})}{k \tan \theta_0}.
\end{equation}
This shows that for given $\theta_0$, the magnitude of $D_{IF}$ is maximum for circularly
polarized \textit{reflected} beams ($\psi= \pi/4$ and $\phi+\Phi_{s0}-\Phi_{p0}= (m+1/2)
\pi$). It also shows that the non-vanishing IF displacement for the case of oblique linear
polarization of the incident beam ($\phi=m \pi$) \cite{Imbert} results from the different
phase shifts between s and p polarizations in total reflection. The incidence angle dependence
$\sim 1/\tan \theta_0$ is in consistency with the recent experimental result
\cite{Pillon-GGLE}. Since $\theta_0$ is larger than the critical angle for total reflection,
it is no wonder that the IF displacement is of the order of $\lambda_0/2 \pi$ \cite{Imbert,
Pillon-GG, Pillon-GGLKG}.

\section{Concluding Remarks}

We have advanced a unified theory for the GH and IF effects by representing the vector angular
spectrum of a 3D light beam in terms of a 2-form angular spectrum consisting of the $s$ and
$p$ polarized components. The 2-form amplitude of the angular spectrum describes the
polarization state of a beam in such a way that the GH displacement is quantized with
eigenstates the 2 orthogonal linear polarizations and the IF displacement is quantized with
eigenstates the 2 orthogonal circular polarizations. We have also derived the Artmann formula
for the GH displacement and found an observable evidence of the so-called translational
inertial spin effect that was discussed more than 40 years ago \cite{Beauregard}. It was shown
that the IF displacement is in fact the translational inertial spin effect happening to the
totally reflected beam.

In the 2-form representation of a bound beam presented here, only large incidence angle
$\theta_0$ in the angular distribution function $A(k_y,k_z)$ corresponds to the uniformly
polarized beams \cite{Pillon-GG, Pillon-GGLKG, Pillon-GGLE}. When $\theta_0$ is very small,
especially when $\theta_0=0$, this representation gives quite different beams with peculiar
polarization distributions, which needs further investigations.

\section*{Acknowledgments}

The author thanks Xi Chen and Qi-Biao Zhu for fruitful discussions. This work was supported in
part by the National Natural Science Foundation of China (Grant 60377025), Science and
Technology Commission of Shanghai Municipal (Grant 04JC14036), and the Shanghai Leading
Academic Discipline Program (T1040).

\end{document}